\documentstyle[12pt]{article}

\renewcommand{\thesubsubsection}{\thesection.\thesubsubsection}
\newcommand{\appsection}{\renewcommand{\thesection}{\Alph{section}}
			\setcounter{section}{0} \setcounter{equation}{0}
			\section{Appendix}}

\font\BB=msym10

\def\Z {\hbox{\BB Z}}

\def\be{\begin{equation}}
\def\ee{\end{equation}}
\def\EQ{\begin{equation}}
\def\EN{\end{equation}}

\def\bea{\begin{eqnarray}}
\def\eea{\end{eqnarray}}
\def\to{\rightarrow}
\def\goto{\longrightarrow}
\def\sa{\hspace{0.1in}}
\def\hs{\hspace{0.1in}}
\def\sb{\hspace{0.2in}}

\def\M{{\cal M}}

\begin{document}
\oddsidemargin 5mm
\setcounter{page}{0}
\renewcommand{\thefootnote}{\fnsymbol{footnote}}
\newpage
\setcounter{page}{0}
\begin{titlepage}
\begin{flushright}
SISSA 192/92/EP
\end{flushright}
\vspace{0.5cm}
\begin{center}
{\large {\bf$S$-Matrices of $\phi_{1,2}$ perturbed unitary
minimal models: IRF-Formulation and Bootstrap-Program
}} \\
\vspace{1.5cm}
{A. Koubek}\\
\vspace{0.8cm}
{\em International School for Advanced Studies, \\
Via Beirut 2-4 \\
34013 Trieste, \\ Italy} \\
\end{center}
\vspace{6mm}
\begin{abstract}
We analyze the algebraic structure of $\phi_{1,2}$ perturbed minimal models
relating them to graph-state models with an underlying Birman-Wenzl-Murakami
algebra. Using this approach one can clarify some physical properties
and  reformulate the bootstrap equations.
These are used to calculate the $S$-matrix elements of higher
kinks, and to determine the breather spectrum of the $\phi_{1,2}$
perturbations of the unitary minimal models $\M_{r,r+1}$.
\end{abstract}
\vspace{5mm}
\end{titlepage}

\newpage

\renewcommand{\thefootnote}{\arabic{footnote}}
\setcounter{footnote}{0}
%\resection{Introduction}
\section{Introduction}
It has been proven by A. Zamolodchikov that certain deformations of
minimal models of conformal field theory (CFT) are described by integrable
massive field theories \cite{ZZ}. The corresponding Hamiltonian can be
written as
\EQ
H_p=H_{CFT} + \lambda \int \hs \phi (x) \hs dx \hs\hs .
\label{ham} \EN
Integrability of the perturbed theory is achieved only for few specific
operators $\phi$ of the space of
states of the CFT, in general only for the primary fields
$\Phi_{1,2}, \hs \Phi_{2,1}$ and $\Phi_{1,3}$ of the Kac-table \cite{ZZ}.

The on-shell behaviour of massive quantum field theories is described by
the $S$-matrix. For integrable massive quantum field theories the
$S$-matrix is factorized, {\em i.e.} $n$-particle scattering amplitudes can be
decomposed into $2$-particle ones.
There is a large variety of methods in order to compute the $S$ matrix
(see {\em e.g.} \cite{crep}), but in many cases the latter is just
conjectured on the basis of physical features and symmetries of the model
under consideration.

We will discuss mainly $\phi_{1,2}$-perturbed minimal models. The
scattering theories corresponding to the hamiltonian
\EQ
H=H_{CFT} + \lambda \int \, \Phi_{1,2} (x) \hs dx\sb ,
\EN
have first been
 discussed by Smirnov \cite{smirnov-12}. He wrote down the $S$-matrix of
the fundamental particle, which in the IRF (Interaction around Face)
representation takes the
form
\begin{eqnarray}
S(\beta) &=& \left(\sinh\frac{\pi}{\xi}(\beta-i\pi)
\sinh\frac{\pi}{\xi}\left(\beta-\frac{2\pi i}{3}\right)\right)^{-1} \nonumber\\
&\times& \exp\left(
-2i \int_0^{\infty} \frac{dx}{x} \frac{\sin \beta x \sinh\frac{\pi
x}{3} \cosh\left(\frac{\pi}{6}-\frac{\xi}{2}\right)x}
{\cosh\frac{\pi x}{2} \sinh\frac{\xi x}{2}}\right) \sa \times R_{bd}^{ac}
\sb.\label{s-smi}
\end{eqnarray}
For our purpose we write the $R$-matrix as
\be
R_{bd}^{ac} =
(-1) \left [
\left \{
\begin{array}{ccc}
1 &b& a\\
1&d&c\end{array}
\right \}_q
\sin \frac{i \pi \beta}\xi
\cos ( \frac{i \pi \beta}\xi +(\alpha+3)\gamma)+
\delta_{ac}\,
\cos 3 \gamma \sin 2 \gamma \right ] \sa ,
\ee
which is equivalent to the expression originally given by Smirnov.
The parameter $\gamma = \frac{r}{s} \pi$ corresponds to the model
$M_{r,s}$  and determines the reduction of the
quantum group, that is neighboring states $a_k,a_{k+1}$ are
restricted by the IRF rules
\EQ
\vert a_k - 1 \vert \, \leq \, a_{k+1} \, \leq \hs
\min \,(a_k + 1, r-3-a_k) \hs . \label{limitation}
\EN
The parameter $\xi=\frac 23 \frac{\pi\gamma}{2 \pi-\gamma}$
relates the rapidity to the spectral parameter of
the quantum group \cite{smirnov-12}.

The spectrum of these theories is rather complicated. This, because the
$S$-matrix also contains poles which generate kinks of higher mass. These
in turn can form further scalar bound-states or even higher kinks. This
mechanism explains the difficulty of finding the spectrum for arbitrary
coupling constant $\gamma$.

In section 2 we analyze the algebraic structure of S-matrices for
$\phi_{1,2}$-perturbed
minimal models, cast them into a graph-state formulation.
In section 3 we confront this construction with the $\phi_{1,2}$ scattering
theories and draw some physical consequences for the ultraviolet limit and
the bootstrap equations.
In section 4 we apply the
bootstrap, find the
$S$-matrices for the higher kink, analyze the pole-structure and conjecture
the full $S$-matrix of all unitary minimal models $M_n$ for $n > 4$
perturbed by $\phi_{1,2}$. Our conclusions are presented in section 5.

\section{Algebraic structure of $\phi_{12}$-perturbed $S$-matrices}

The approach of Smirnov was to use the vector-representation of the
$R$-matrix of $A_2^{(2)}$ and then to perform a change to the
IRF-representation.
This approach has the disadvantage that it passes through the
vector-representation of $A_2^{(2)}$, which gives an inconsistent
field theoretic model, since the corresponding hamiltonian is not
hermitian. In order to construct a consistent field theory, one needs to
restrict the Hilbert space, {\em i.e.} one has to go into the IRF-
representation.

Our goal is to construct the IRF-amplitudes directly. We want to
emphasize, that we will not derive {\em new} models with respect to those of
Smirnov, but our construction will allow us to understand better the
algebraic structure of the scattering amplitudes. We will need this
formulation to find new physical results in section 3.

\subsection{Temperly Lieb Algebra}
The construction is based on models which intrinsically present the
restriction of the Hilbert space: the so-called graph-state models
\cite{japan,france, kuniba}.
These graphs are usually picturized fusion algebras of some
Wess-Zumino-Witten (WZW) model.
For our purpose, describing perturbations of conformal
field theory, we use graphs based on the fusion-rules of $SU(2)$
WZW-models, which read as
\be
\phi_{j_1} \times \phi_{j_2} =
\sum_{j_3=\vert j_1-j_2\vert}^{\min (j_1+j_2,k-j_1-j_2)} \phi_{j_3}
\sb .\ee
For the fundamental representation (spin j=$\frac 12$) they coincide with the
graph of the $A_{k+1}$ Dynkin-Diagrams, where $k$ denotes the level of the
underlying Kac-Moody algebra,
\be
\begin{picture}(400,20)(-70,0)
\put(0,0){\line(1,0){210}}
\put(250,0){\line(1,0){20}}
\put(0,0){\circle*{5}}
\put(0,-10){\makebox(0,0){0}}
\put(40,0){\circle*{5}}
\put(40,-10){\makebox(0,0){$\frac 12$}}
\put(80,0){\circle*{5}}
\put(80,-10){\makebox(0,0){1}}
\put(120,0){\circle*{5}}
\put(120,-10){\makebox(0,0){$\frac 32$}}
\put(160,0){\circle*{5}}
\put(160,-10){\makebox(0,0){2}}
\put(200,0){\circle*{5}}
\put(200,-10){\makebox(0,0){$\frac 52$}}
\put(280,0){\circle*{5}}
\put(280,-10){\makebox(0,0){$\frac k2$}}
\put(230,0){\makebox(0,0){$\dots$}}
\end{picture}
\label{13}
\ee
In order to construct $\phi_{1,2}$-perturbed  models, we are interested in
the spin $j=1$ representation. The corresponding graphs are

\be
\begin{picture}(400,30)(0,-30)
\put(0,0){
 \begin{picture}(60,30)
\thicklines
\put(0,0){\circle*{5}}
\put(0,10){\makebox(0,0){0}}
\put(10,-20){\makebox(0,0){$k=3$}}
\put(0,0){\line(1,0){50}}
\put(50,0){\circle*{5}}
\put(50,10){\makebox(0,0){1}}
\put(50,0){\line(-1,-3){10}}
\put(50,0){\line(1,-3){10}}
\put(50,-30){\oval(20,20)[b]}
\put(60,0){,}
\end{picture}
}
\put(90,0){
 \begin{picture}(110,30)
\thicklines
\put(0,0){\circle*{5}}
\put(0,10){\makebox(0,0){0}}
\put(10,-20){\makebox(0,0){$k=4$}}
\put(0,0){\line(1,0){50}}
\put(50,0){\circle*{5}}
\put(50,10){\makebox(0,0){1}}
\put(50,0){\line(-1,-3){10}}
\put(50,0){\line(1,-3){10}}
\put(50,-30){\oval(20,20)[b]}
\put(50,0){\line(1,0){50}}
\put(100,0){\circle*{5}}
\put(100,10){\makebox(0,0){2}}
\put(110,0){,}
\end{picture}
}
\put(230,0){
 \begin{picture}(110,30)
\thicklines
\put(0,0){\circle*{5}}
\put(0,10){\makebox(0,0){0}}
\put(10,-20){\makebox(0,0){$k=5$}}
\put(0,0){\line(1,0){50}}
\put(50,0){\circle*{5}}
\put(50,10){\makebox(0,0){1}}
\put(50,0){\line(-1,-3){10}}
\put(50,0){\line(1,-3){10}}
\put(50,-30){\oval(20,20)[b]}
\put(50,0){\line(1,0){50}}
\put(100,0){\circle*{5}}
\put(100,10){\makebox(0,0){2}}
\put(100,0){\line(-1,-3){10}}
\put(100,0){\line(1,-3){10}}
\put(100,-30){\oval(20,20)[b]}
\put(110,0){,}
\put(120,0){\dots}
\put(142,0){,}
\end{picture}  }
\end{picture}
\label{12} \ee
where we indicated also the corresponding level of the Kac-Moody algebra.
Given a graph, one can find a representation of the Temperly-Lieb algebra
(TLA),
\bea
& &E_i E_j = E_j E_i\sa\sa {\rm for} \sa \vert i-j\vert \ge 2\sa,\\
& &E_iE_{i\pm 1}E_i =E_i\sa, \sb E_i^2 =\sigma(j)^{\frac 12} E_i\sa,
\eea
by diagonalizing the incidence matrix of the diagram.
The generators $E_i$ operate in an $n$-particle space. For graph-state
models this action can be visualized as
\be
\begin{picture}(270,50)(0,-25)
\put(0,0){\line(0,-1){50}}
\put(50,0){\line(0,-1){50}}
\put(200,0){\line(0,-1){50}}
\put(250,0){\line(0,-1){50}}
\put(100,0){\line(1,-1){50}}
\put(150,0){\line(-1,-1){50}}
\put(25,-25){\makebox(0,0){$l-2$}}
\put(75,-25){\makebox(0,0){$l-1$}}
\put(175,-25){\makebox(0,0){$l+1$}}
\put(225,-25){\makebox(0,0){$l+2$}}
\put(125,-40){\makebox(0,0){$l$}}
\put(125,-10){\makebox(0,0){$l'$}}
\put(-20,-25){$\dots$}
\put(258,-25){$\dots$}
\end{picture}
\sim
E_{l-1,l+1}^{l,l'} \sb ,
\label{tla-graph}
\ee
\vspace{5mm}

\noindent
and the generators are matrices in the indices $l$ and $l'$.
It is amazing how similar the results are for the two different families of
graphs (\ref{13}) and (\ref{12}).
Let us define the parameter $\lambda \equiv \frac\pi {k+2}$. Then the largest
Eigenvalue, corresponding to the Perron-Frobenius Eigenvector\footnote{ The
other eigenvalues lead in general to imaginary Boltzmann weights, for
an example see \cite{a6} }
for the case (\ref{13}) is $\sigma(\frac 12)=2 \cos \lambda$ whereas for
the case (\ref{12}) it is $\sigma(1)=1+2 \cos 2 \lambda$.
This becomes more
similar introducing a notion borrowed from the quantum-group language.
Let us define $q=e^{i \lambda}$ and the quantum-symbol
$[n] = \frac {q^n-q^{-n}}{q-q^{-1}}$. Then we find that
$\sigma(\frac 12) = [\frac 12]$ and $\sigma(1) = [1]$.

Also the
eigenvectors have the same structure. They are $\psi(a) = [2 a+1]$, where
the numbers $a$ take the values of the labels on the nodes of the
corresponding diagram, that is half-integers for (\ref{13}) and integers
for (\ref{12}) respectively. These indices $a$ are restricted in both
cases by the bound $ a \le \frac k2$.
Finally, the generators (\ref{tla-graph}) are constructed out of the
eigenvectors \cite{japan} as
\be
E_{bd}^{ac} = \frac{[2a+1]^{\frac 12}[2c+1]^{\frac 12}}{[2b+1]^{\frac 12}
[2d+1]^{\frac 12}} \delta_{bd}
\sb .\ee

\subsection{Braid Group}
In order to introduce the spectral parameter one needs to go to a
braid-group representation, that is elements $b_i$ satisfying
\bea
& &b_ib_j=b_jb_i\sa, {\rm for}\sa \vert i-j\vert \ge 2 \sa ,\nonumber \\
& &b_ib_{i+1}b_i = b_{i+1}b_ib_{i+1} \sb .
\eea
The usual approach, which leads to $\phi_{1,3}$ perturbed models,
defines the braid-group generators by the linear transformation
\be b_i=1-e^{i \lambda} E_i\sb .
\ee
In that way one obtains a Hecke algebra, since the linear transformation
supplies a quadratic relation for the Braid group generators,
\be (b_i-1)(b_i+e^{2 i \lambda})=0 \sb .
\label{2order}
\ee

For the spin-1 algebra the natural choice \cite{japan}
 is the Birman-Wenzl-Murakami (BWM) algebra \cite{bwm}, which
is given by the relations
\be
\begin{array}{l}
g_ig_j=g_jg_i\sa,\sa {\rm for}\sa \vert i-j\vert \ge 2\sa, \\
g_ig_{i+1}g_i = g_{i+1}g_ig_{i+1} \sa;\\
\\
e_i e_j = e_j e_i\sa, \sa{\rm for} \sa \vert i-j\vert \ge 2\sa,\\
e_ie_{i\pm 1}e_i =e_i\sa, \sb e_i^2 =(m^{-1} (l+l^{-1})-1) e_i \sa;\\
\\
g_i+g_i^{-1} = m(1+e_i) \sa,\sa g_i^2 = m (g_i +l^{-1} e_i ) -1 \sa,\\
g_{i+1}g_i e_{i\pm 1} = e_i g_{i\pm 1}g_i = e_ie_{i\pm 1} \sa,\\
g_{i\pm1}e_ig_{i\pm1} = g_i^{-1} e_{i\pm 1}g_i^{-1} \sa,\sa
e_{i\pm1}e_ie_{i\pm1} = g_i^{-1} e_{i\pm 1}\sa,\\
e_{i\pm1}e_ig_{i\pm1} = e_{i\pm 1}g_i^{-1} \sa,\sa
g_ie_i=e_ig_i =l^{-1}e_i \sa,\sa e_ig_{i\pm1}e_i =l e_i
\end{array}
\label{BWM}
\ee
with $e_i = - E_i$ and $g_i = -i b_i$. The parameters appearing in the
algebra are  $m=-i (q^2-q^{-2})$ and $l=i q^4$.
This algebra implies a third order relation for the braid group
generators \cite{japan},
which in our notation reads as
\be (b_i-q^{-2})(b_i+q^2)(b_i+q^{-4})=0\sb.
\label{3order}
\ee
Not all of the relations in (\ref{BWM}) are independent \cite{li-wang}, but
in order to clearly see the relation of braid group and Temperly-Lieb
algebra we listed them anyway.

One can again find
generators $b_i$ satisfying (\ref{BWM}) which are of  the same form
as those satisfying (\ref{2order}) for the corresponding spin,
 if we write them in quantum group language. For that it is necessary to
introduce the $6j$-symbols \cite{kir-res},
\be \begin{array}{c}
\left \{ \begin{array}{ccc} a & b & e \\ d
&c & f  \end{array} \right \}_q
= \sqrt{ [2 e +1 ] [ 2 f + 1]} \sa (-1)^{c+d+2e-a-b} \sa \times
\\
  \Delta (abe) \Delta (acf) \Delta (dce) \Delta (dbf)
\sa \sum _{z} (-1)^z [z+1]! \sa \times \\
  \left ( [z-a-b-e]! [z-a-c-f]!
[z-b-d-f]![z-d-c-e]! \right .  \sa \times \\
  \left . [a+b+c+d-z]! [a+d+e+f-z]! [b+c+e+f-z]! \right ) ^{-1} \sb ,
\end{array} \label{6jsym} \ee
wherein we use the conventions that $[0]!=1$ and the sum
runs only over $z$ such that no factor $[x]$ is less than zero.
Further,
$$ \Delta(abc) = \left ( \frac {
[ -a+b+c]![a-b+c]![a+b-c]!}{[a+b+c+1]!} \right )^{\frac 12} \sa .$$
With this definition the braid group generators can be written as
\be
b_{bd}^{ac}=   q^{(c_d-c_c-c_a+c_b)}
(-1)^{(b+d-a-c)} \times
\left \{ \begin{array}{ccc} j & b & a \\ j
&d & c  \end{array} \right \}_q \times (-1)^{(a-c)\frac 1j} \sb .
\label{b-a22}\ee
These generators satisfy a further property: the crossing symmetry,
\be
b_{bd}^{ac} = \left ( b_{ac}^{bd} \right )^{-1} \left (
\frac{[2 a+1][2c+1]}{[2b+1][2d+1]}\right )^{\frac 12}\sb.
\label{b-cross}
\ee

\subsection{Introducing the Spectral Parameter}
In \cite{jones} it is shown that given a representation of the braid
group
which factors either through the Hecke algebra or through the BWM algebra,
one can introduce a spectral parameter with a mechanism called
{\em universal baxterization}.
In the Hecke algebra case one finds
\be R_i(x)=q^{-1} e^x b_i - q e^{-x} b_i^{-1} \sb ,
\label{r-a11}\ee
whereas for the BWM case the spectral parameter depending solution
is written as
\be
R_i(x)=(x^{-1}-1)k g_i + m (k+k^{-1}) + (x-1)k^{-1} g_i^{-1}\sb,
\label{r-a22}
\ee
with $k=q^3$.
Using the BWM-algebra (\ref{BWM}) and the relation (\ref{b-cross})
one can show, that the $R$-matrix (\ref{r-a22}) satisfies crossing
\be R_{bd}^{ac}(x) = \left (
\frac{[2 a+1][2c+1]}{[2b+1][2d+1]}\right )^{\frac 12}
R_{ac}^{bd} (-x^{-1}q^6)\sb ,
\label{r-cross}
\ee
and the completeness relation
\bea & & \sum_{e} R_{bd}^{ae} (x) R_{bd}^{ec} (-x) = \nonumber \\
 &=& \delta _{ac} \, \times
(x^{-1}q^3 +x q^{-3})(x^{-1}q^2-x q^{-2})(x q^3+x^{-1}q^{-3})(
x q^{-2}-x^{-1}q^2) \sa .
\label{comp}
\eea
Similar results are well known for the $R$-matrix based on the
Hecke algebra (\ref{r-a11}) \cite{japan}.

As a last note we mention the so-called {\em symmetry-breaking
transformations} \cite{japan}, which leave untouched the Yang-Baxter
equation and the
 completeness-relation, but can change the parameters appearing in the
crossing-relation. The transformations we will need are
\be
R_{bd}^{ac}(e^u) \to
{\tilde R}_{bd}^{ac}(e^u) =
\alpha_{abcd}(u) . \beta_{abcd} \times R_{bd}^{ac} (e^u)\sb ,
\ee
with
\bea
\alpha_{abcd}(u)&=& e^{[-p(a)+p(b)-p(c)+p(d)]u} \sa ,\label{trans1}\\
\beta_{abcd} &=& \frac{p'(a)}{p'(c)}\label{trans2}\sb.
\eea
Herein $p(.)$ and $p'(.)$ are arbitrary functions. The
transformation (\ref{trans1}) can be used to eliminate the parameters
appearing in the crossing relation \cite{leclair,CKM2}. The second
transformation is of particular importance for relating the above
$R$-matrix to the Izergin-Korepin $R$-matrix used by Smirnov.
In order to transform (\ref{r-a22}) to become equal to
(\ref{s-smi}), we need to perform a gauge-transformation
of the form $R_{bd}^{ac}(x) \to (-1)^{a-c} R_{bd}^{ac}(x)$.
This is of physical significance because it changes the signs of some
amplitudes.
Since the signs of the residues in a unitary theory are
 fixed, this simple gauge-transformation can not be obscured. Anyhow,
note that with this gauge transformation also the underlying braid-group
and TLA undergo the same transformation. To simplify the discussion, we
consider from now on the $R$-matrix (\ref{r-a22}), inserting the factor
$(-1)^{a-c}$, leaving though the form with non-trivial
crossing factors.

\section{Application to Scattering Theories}

Now we want to apply this mathematical construction to the problem of
scattering theories describing deformations of conformal field theories.
{}From now on, we will concentrate our discussion mainly onto the $R$-matrix
built on the BWM-algebra, since this is the one describing $\phi_{1,2}$
perturbed models, which we are mainly interested in. Analogous results hold
for the $R$-matrix constructed from the Hecke algebra describing
$\phi_{1,3}$ perturbed models.
These theories have been analyzed in
\cite{leclair,smirnov-13}.

\subsection{Construction of the Scattering Amplitudes}
In order to identify the corresponding scattering theory one needs to relate
the
spectral parameter $x$ to the rapidity variable. This causes that a whole
series of scattering theories get related to the same $R$-matrix.
Let us explain this mechanism for the $R$-matrix (\ref{r-a22}).

Since we have in mind the scattering theories proposed by Smirnov, we
make an ansatz, $x=e^\frac{2 \pi \beta}{\xi}$. Crossing symmetry in
scattering theories means $S(\beta)_{bd}^{ac} = S(i\pi-\beta)_{ac}^{bd}$.
For the $R$-matrix we have the relation (\ref{r-cross}), which includes
also the factors arising from a symmetry-breaking transformation.
In order to achive crossing symmetry the transformation
$\beta \to i\pi-\beta$ must correspond to
$x\to - x^{-1} q^6$. This implies the constraint
\be
\frac{2 \pi^2 i}{\xi} = i\pi +\frac{6 i\pi}{r}+2 n \pi i\sb,
\ee
from which  we find the relation for the parameter $\xi=
\frac{2 \pi r}{\pm 6+2 n r - 3 r}$, with $n\in \Z$ and with $r\equiv k+2$.
 But in order to implement
the symmetry of the diagram (\ref{12}) dynamically we need a bound state
at the pole $\beta = \frac{2 i \pi}{3}$. But this requires that at
this point the $R$-matrix has to degenerate into a 3-dimensional projector.
In the appendix we have collected some information on the projectors and
the necessary $6j$-symbols.
Therefore we need that
$$\left .e^{\frac{2 \pi \beta}{\xi}}\right \vert_{\beta=\frac{2\pi i}{3}}
=q^4 \sb .$$
This condition eliminates part of the possible values for the parameter
$\xi$, leaving \linebreak
$\xi=\frac{2 \pi r}{\pm 6+6 n r - 3 r}$.

This condition is equivalent to an idea raised by Zamolodchikov
\cite{zam-sub}, constructing a
factorizable scattering theory for the tricritical Ising model perturbed
by the subleading magnetization ( $\M_{4,5}+\phi_{2,1}$ ). He required
that one of the amplitudes needs to become zero at the pole at $\frac{2
\pi i}{3}$. The amplitude corresponds to $R_{00}^{11}$, since a kink
interpolating the vacuum $0$ to the vacuum $0$ does not exist
(see the graph (\ref{12}), which has no tadpole at the node $0$). This
condition is automatically fulfilled if the amplitudes degenerates into a
three-dimensional projector at this point.

To show this we need the formulation of the previous paragraph of the
$R$-matrix in terms of $6j$-symbols. Since this $R$-matrix is an
affinization of a quantum group in the shadow-world representation
\cite{kir-res}, we can also express the projectors as $6j$-symbols, that is
\be
P_{bd}^{ac;j}=
\left \{ \begin{array}{ccc}
1& 1& j\\
b&d&a
\end{array}
\right \}
\left \{ \begin{array}{ccc}
1& 1& j\\
b&d&c
\end{array}
\right \}\sb .
\label{projectors}
\ee
The exact relation for the $3d$-projector is
\be R_{bd}^{ac} (x=q^4,q) = [2][4]
P_{bd}^{ac;1}\sb .
\ee
{}From the expressions given in the appendix, we easily can calculate the
residues at the pole. The general amplitude needed in order to verify the
Zamolodchikov condition is
$$R_{ll}^{l+1,l+1}(q^4,q) = (q^2-q^{-2})[2] \frac{[2l][2l+3]}
{[2l+2][2l+1]}\sb .
$$
This becomes zero for $l=0$.

As a last ingredient for a physical scattering theory one needs unitarity,
that is $S(\beta) S^{\star}(\beta)=1$.
Since the elements of the $R$-matrix are real\footnote{This is
related to the fact, that we
used the highest eigenvalue in diagonalizing the incidence matrix of the
diagrams (\ref{12}), whose corresponding
eigenvector is the Frobenius-Perron eigenvector.}, $R$ satisfies also
real analyticity, {\em i.e.} $S^{\star}(\beta) = S(-\beta)$.
Additionally we have the completeness property
(\ref{comp}), and therefore the $R$-matrix multiplied  by a scalar factor
$S_0$, which eliminates the terms on the right hand side of
(\ref{comp}) is unitary.
But this factor has been determined by Smirnov, and is the prefactor in
(\ref{s-smi}) with the corresponding parameter $\xi$.

Confronting the resulting theories with those of Smirnov (\ref{s-smi}), we find
that all of the perturbed conformal scattering theories $\M_{r,m r\pm 1}
+\phi_{1,2}$ correspond to the $R$-matrix (\ref{r-a22}). Here $m$ takes the
values $ m=1,2,\dots$. Of course the `formal' theory $\M_{r,r-1} +\phi_{1,2}$
is the physical one $\M_{r-1,r} +\phi_{2,1}$.
For all of these theories the scattering matrix of the fundamental particle
is unitary, that is $S S^\star =1$. Since through the bootstrap this
property is preserved also for other particles, {\em all} of these models are
supposed to be consistent scattering theories.

We want to insert a comment here. We found that {\em one} $R$-matrix
corresponds to {\em many} different scattering theories, according to how
one relates the rapidity variable to the spectral parameter.
 Up to now,
there was the believe that there is a unique way to find a physical
scattering theory given an $R$-matrix, using the principle of ``minimality''.
This principle was commonly used in order to eliminate ambiguities deriving
from the fact that the factor $S_0$ can not be derived uniquely, but has
always an ambiguity of so-called CDD-factors. Minimality said, that the
physical scattering theory corresponding to a given $R$-matrix is that one,
which introduces the minimal number of poles and zeros in the physical
strip. We see now, that this is no fundamental principle. We find, that
the theories belonging to one $R$-matrix depend on how the spectral
parameter is related to the rapidity variable, and the $S$-matrix of the
fundamental particle differ from each other by CDD-factors.
These factors of course usually introduce further poles in the physical
strip, and therefore generate a completely different physical scattering
theory. This fact was explicitly discussed for scattering theories of
perturbed minimal models $\M_{5,n}$ in \cite{a6}.

Analyzing the allowed parameters $\xi$ we find that the theory with the minimal
number of
poles and zero's is that one, corresponding to a deformed {\em unitary}
conformal theory. But note that also the $S$-matrices of the fundamental
particles of $\M_{r-1,r}+\phi_{2,1}$ and the theory $\M_{r,r+1}$
correspond to the same $R$-matrix $R(x)$.

\subsection{Ultraviolet Limit}
As a next point, let us discuss the ultraviolet limit. For $\beta \to
\infty$ the $S$-matrix becomes again proportional to the braid-group
generators (\ref{b-a22}), but with the gauge-transformation, that is
\be
b_{bd}^{ac}= S_0 (\infty) \, q^{(c_d-c_c-c_a+c_b)}
(-1)^{(b+d-a-c)} \times
\left \{ \begin{array}{ccc} j & b & a \\ j
&d & c  \end{array} \right \}_q  \sb .
\label{b-new}\ee
This expression is valid also for $\phi_{1,3}$ perturbations, which
correspond to the spin $j=\frac 12$. One notices that these expressions are
proportional to the braiding matrices of conformal blocks of the WZW-models
\cite{alv}, as one expects.

Now we use the algebraic structure. Let us view these braid-group
generators as matrices in the indices $a$ and $c$. Now since they satisfy
(for spin $j=1$) a third order relation (\ref{3order}), there can only be 3
 independent eigenvalues. The same fact holds also for
the corresponding $R$-matrices, whose non-diagonal components are given by
the braid group generators. Diagonalizing those one finds that the
eigenvalues correspond to the amplitudes
$S_{00}^{11}(\beta)$,
$S_{01}^{11}(\beta)$ and
$S_{02}^{11}(\beta)$ which define three independent
phase-shifts.
We study now their asymptotic behaviour
\begin{eqnarray}
\lim_{\beta\rightarrow \infty} \hs S_{00}^{11}(\beta) & = &
e^{2 i \pi \Delta_{3,1} } \hs,\nonumber \\
\lim_{\beta\rightarrow\infty} \hs S_{01}^{11}(\beta) & = &
e^{ i \pi \Delta_{3,1} } \hs ,\label{asympbeh}
\\
\lim_{\beta\rightarrow\infty} \hs S_{00}^{11}(\beta) & = &
e^{ i \pi (2\Delta_{3,1} - \Delta_{5,1}) } \hs,\nonumber
\end{eqnarray}
where $\Delta_{3,1}$ and $\Delta_{5,1}$ are the anomalous dimensions
of the corresponding fields of the original conformal field theory.
These are exactly the dimensions appearing in the
operator product expansion (OPE) of $\Psi\equiv\Phi_{3,1}$ of the original
minimal model ${\cal M}_{r,mr\pm 1}$:
\EQ
\Psi(z) \Psi(0) = \frac{1}{z^{2\Delta_{3,1}}} \hs {\bf 1} +
\frac{C_{\Psi,\Psi,\Psi}}{z^{\Delta_{3,1}}} \hs \Psi(0) +
\frac{C_{\Psi,\Psi,\Phi_{5,1}}}{z^{2\Delta_{3,1}-\Delta_{5,1}}} \hs
\Phi_{5,1}(0) + \ldots
\EN
Of course if one considers the series of theories $\M_{r-1,r}+\phi_{2,1}$
one finds that the corresponding field is $\phi_{1,3}$ instead of
$\phi_{3,1}$. This correspondence for $\phi_{2,1}$ perturbed unitary
theories has been found in \cite{CKM2}.

Similar one can analyze the asymptotic phase-shifts of the $\phi_{1,3}$-
perturbed models. They satisfy a second order relation (\ref{2order}) and
therefore the
braid group generators as well as the $R$-matrix (\ref{r-a11}) have
only two eigenvalues. They correspond to the amplitudes
$S_{00}^{\frac 12 \frac 12}(\beta)$ and
$S^{\frac 12 \frac 12}_{01}(\beta)$.
It is not surprising that their asymptotic phase-shifts determine the
dimensions of the OPE of the field $\phi_{2,1}$,
\be
S_{00}^{\frac 12 \frac 12}(\infty)=e^{2 i \pi \Delta_{2,1}}\sb ,
S^{\frac 12 \frac 12}_{01}(\beta)=e^{i\pi(2\Delta_{2,1}-\Delta_{3,1})}
\sb .
\ee

\subsection{Bootstrap Equations}
The last formal application involves the bootstrap-equations.
For degenerate particles in the IRF description
they were developed in \cite{chim-zam}. The equations are
\be
S_{ad}^{bd} f_{abc} =
\sum_{g} f_{egd} S_{ag}^{eb} (\theta+i \bar{u} ) \, S_{db}^{cg}
(\theta-i\bar{u}) \sb .
\ee
The relation of the constants $f$ with the scattering matrix
\cite{KM} is
\be
{\rm Res}_{\theta = i u}
S_{bd}^{ac} =
i f_{bad} f_{bcd}\sb ,
\ee
where $u$ is the corresponding $S$-matrix pole.
It is useful to exploit the quantum-group symmetry in order to reformulate
the above equations, since the above definition of the constants $f$ leads
to a system of quadratic equations to solve and therefore leaves an
ambiguity of a sign.

Since the $S$-matrix for $\phi_{12}$ perturbed minimal models is
proportional to the $A_2^{(2)}$ quantum group $R$-matrix, one can also
rewrite the bootstrap-equations in terms of the pentagon-identity.
This determines the constants $f$ as $6j$-symbols. Or more explicitly,
\be
f_{abc} =
\left \{ \begin{array}{ccc}
1& 1& j\\
a&c&b
\end{array}
\right \}\sb ,
\ee
where the spin $j$ corresponds to the projector, into which the $S$-matrix
degenerates at the pole. This correspondance can also be seen from the
form of the projectors  (\ref{projectors}).

\section{Bootstrap for the unitary series}

The unitary minimal series perturbed by the operator $\phi_{12}$ was
analyzed by Smirnov\cite{smirnov-12}.
 He established the spectrum of all theories except
$M_{5,6}$ and wrote down the $S$ matrix of the fundamental
kink as well as that one of the fundamental breather. We apply now the
bootstrap-equations in the IRF formulation in order to write down the
complete $S$-matrix of kinks and their bound-states, the breathers.
As a byproduct we also find that the theory $M_{5,6}$ has
two kinks and four breathers.

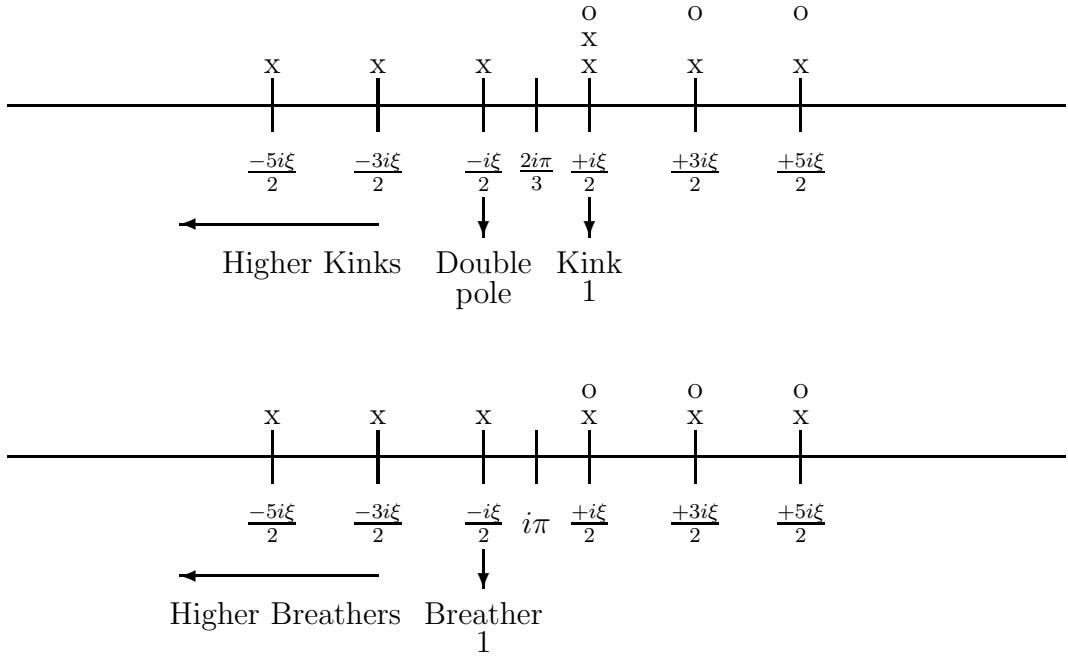
\begin{figure}
$$
 \begin{picture}(450,120)(0,-100)
\thicklines
\put(0,0){\line(1,0){400}}
\put(200,10){\line(0,-1){20}}
\put(200,-25){\makebox(0,0){$\frac{2 i \pi}3$}}
\multiput(100,10)(40,0){6}{\line(0,-1){20}}
\multiput(100,15)(40,0){6}{\makebox(0,0){x}}
\put(220,25){\makebox(0,0){x}}
\put(220,35){\makebox(0,0){o}}
\put(260,35){\makebox(0,0){o}}
\put(300,35){\makebox(0,0){o}}
\put(100,-25){\makebox(0,0){$\frac{-5 i \xi}2$}}
\put(140,-25){\makebox(0,0){$\frac{-3 i \xi}2$}}
\put(180,-25){\makebox(0,0){$\frac{- i \xi}2$}}
\put(220,-25){\makebox(0,0){$\frac{+i \xi}2$}}
\put(260,-25){\makebox(0,0){$\frac{+3 i \xi}2$}}
\put(300,-25){\makebox(0,0){$\frac{+5 i \xi}2$}}
\put(180,-35){\vector(0,-1){15}}
\put(220,-35){\vector(0,-1){15}}
\put(180,-55){\makebox(0,0)[t]{\shortstack{Double\\pole}}}
\put(220,-55){\makebox(0,0)[t]{\shortstack{Kink\\1}}}
\put(140,-45){\vector(-1,0){75}}
\put(115,-55){\makebox(0,0)[t]{Higher Kinks}}
\end{picture}
$$
$$
 \begin{picture}(450,120)(0,-100)
\thicklines
\put(0,0){\line(1,0){400}}
\put(200,10){\line(0,-1){20}}
\put(200,-25){\makebox(0,0){$ i \pi$}}
\multiput(100,10)(40,0){6}{\line(0,-1){20}}
\multiput(100,15)(40,0){6}{\makebox(0,0){x}}
\put(220,25){\makebox(0,0){o}}
\put(260,25){\makebox(0,0){o}}
\put(300,25){\makebox(0,0){o}}
\put(100,-25){\makebox(0,0){$\frac{-5 i \xi}2$}}
\put(140,-25){\makebox(0,0){$\frac{-3 i \xi}2$}}
\put(180,-25){\makebox(0,0){$\frac{- i \xi}2$}}
\put(220,-25){\makebox(0,0){$\frac{+i \xi}2$}}
\put(260,-25){\makebox(0,0){$\frac{+3 i \xi}2$}}
\put(300,-25){\makebox(0,0){$\frac{+5 i \xi}2$}}
\put(180,-35){\vector(0,-1){15}}
\put(180,-55){\makebox(0,0)[t]{\shortstack{Breather\\1}}}
\put(140,-45){\vector(-1,0){75}}
\put(105,-55){\makebox(0,0)[t]{Higher Breathers}}
\end{picture}
$$
\caption{Pole structure of the $S$-matrix $S_{K_1,K_2}$}
\label{fig-12}
\end{figure}

\begin{figure}
\begin{center}
\begin{picture}(145,130)
\thicklines
\put(50,50){\line(-1,-1){20}}
\put(50,50){\line(0,1){30}}
\put(50,80){\line(-1,1){20}}
\put(95,50){\line(1,-1){20}}
\put(95,50){\line(0,1){30}}
\put(95,80){\line(1,1){20}}
\put(50,50){\line(3,2){45}}
\put(50,80){\line(3,-2){45}}
\put(20,20){\makebox(0,0){$K_1$}}
\put(125,110){\makebox(0,0){$K_1$}}
\put(125,20){\makebox(0,0){$K_2$}}
\put(20,110){\makebox(0,0){$K_2$}}
\put(40,65){\makebox(0,0){$K_1$}}
\put(105,65){\makebox(0,0){$K_1$}}
\put(60,44){\makebox(0,0)[b]{$K_1$}}
\put(85,45){\makebox(0,0)[b]{$K_1$}}
\put(83,88){\makebox(0,0)[t]{$K_1$}}
\put(62,86){\makebox(0,0)[t]{$K_1$}}
\put(72.5,65){\circle*{8}}
\end{picture}
\end{center}

\begin{center}
\begin{picture}(145,130)
\thicklines
\put(50,50){\line(-1,-1){20}}
\put(50,50){\line(0,1){30}}
\put(50,80){\line(-1,1){20}}
\put(95,50){\line(1,-1){20}}
\put(95,50){\line(0,1){30}}
\put(95,80){\line(1,1){20}}
\put(50,50){\line(3,2){45}}
\put(50,80){\line(3,-2){45}}
\put(20,20){\makebox(0,0){$K_2$}}
\put(125,110){\makebox(0,0){$K_2$}}
\put(125,20){\makebox(0,0){$K_2$}}
\put(20,110){\makebox(0,0){$K_2$}}
\put(40,65){\makebox(0,0){$K_1$}}
\put(105,65){\makebox(0,0){$K_1$}}
\put(60,44){\makebox(0,0)[b]{$K_1$}}
\put(85,45){\makebox(0,0)[b]{$K_1$}}
\put(83,88){\makebox(0,0)[t]{$K_1$}}
\put(62,86){\makebox(0,0)[t]{$K_1$}}
\put(72.5,65){\circle*{8}}
\end{picture}
\begin{picture}(145,130)
\thicklines
\put(50,50){\line(-1,-1){20}}
\put(50,50){\line(0,1){30}}
\put(50,80){\line(-1,1){20}}
\put(95,50){\line(1,-1){20}}
\put(95,50){\line(0,1){30}}
\put(95,80){\line(1,1){20}}
\put(50,50){\line(3,2){45}}
\put(50,80){\line(3,-2){45}}
\put(20,20){\makebox(0,0){$K_2$}}
\put(125,110){\makebox(0,0){$K_2$}}
\put(125,20){\makebox(0,0){$K_2$}}
\put(20,110){\makebox(0,0){$K_2$}}
\put(40,65){\makebox(0,0){$K_1$}}
\put(105,65){\makebox(0,0){$K_1$}}
\put(60,44){\makebox(0,0)[b]{$B_1$}}
\put(85,45){\makebox(0,0)[b]{$B_1$}}
\put(83,88){\makebox(0,0)[t]{$B_1$}}
\put(62,86){\makebox(0,0)[t]{$B_1$}}
\put(72.5,65){\circle*{8}}
\end{picture}
\end{center}
\caption{Multi scattering processes responsible for higher order poles
in the $S$-matrices  $S_{K_1,K_2}$ and $S_{K_2,K_2}$.}
\label{fig-double}
\end{figure}
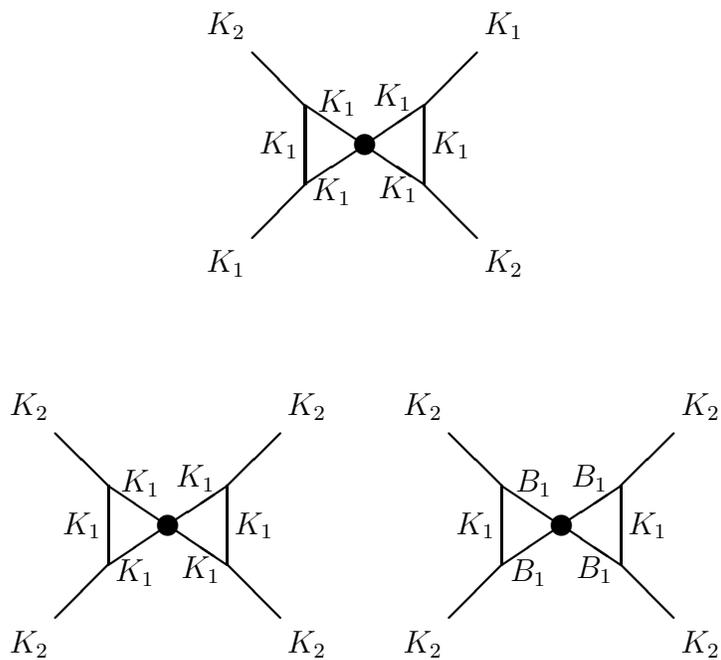

The calculation of the $S$-matrices is tedious, but straightforward.
For that we give only the results.
Let us use the abbreviations
\be
(x)^\pm
= \frac{\Gamma ( \frac{2 k \pi}\xi + x \pm \frac{i \beta}\xi)}
{\Gamma ( \frac{2 k \pi}\xi + x \mp \frac{i \beta}\xi)}\sb,
\ee
and
\be \langle x \rangle =\frac { \tanh ( \frac{\theta}2 + i \pi x)}{
\tanh ( \frac{\theta}2 - i \pi x)}
\sb . \label{scalar-s} \ee
Then the $S$-matrices of the kinks are
\bea& & S_{K_1,K_1}(\beta) =
\left ( \sinh \frac \pi\xi (\beta-i \pi)\, \sinh \frac \pi\xi ( \beta -
\frac{2 \pi i}{3} )\right )^{-1}\,\times
\nonumber \\
& &
\prod_{k=0}^{\infty}
(\frac \pi\xi)^-(\frac{2 \pi}\xi)^+(1)^+(1+\frac\pi\xi)^-
(\frac\pi{3\xi})^+(\frac{4\pi}{3\xi})^-
(1+\frac{2\pi}{3\xi})^-(1+\frac{5\pi}{3\xi})^+
\times
R_{bd}^{ac}
\eea
\bea S_{K_1,K_2}(\beta) &=&
\left ( \cosh \frac \pi\xi (\beta-i \pi) \cosh \frac \pi\xi ( \beta -
\frac{2 \pi i}{3} )\right )^{-1}
\nonumber \\
& &
\prod_{k=0}^{\infty}
(\frac 12+\frac{\pi}{\xi})^-
(\frac 12+\frac{2 \pi}{\xi})^+
(\frac 12)^+
(\frac 12+\frac{\pi}{\xi})^- \times\nonumber \\
& &
(\frac 12+\frac{\pi}{3\xi})^+
(\frac 12+\frac{4\pi}{3\xi})^-
(\frac 12+\frac{2\pi}{3\xi})^-
(\frac 12+\frac{5\pi}{3\xi})^+ \times\nonumber\\
& &
\tilde{R}_{bd}^{ac}  \, \times
\langle \frac{2 i \pi}3 -\frac\xi 2\rangle \langle \frac{2\pi}3+\frac\xi
2\rangle
\sb ,\eea
where $\tilde{R}$ is the $R$-matrix with a spectral parameter
shifted by a phase-factor of $\frac\pi 2$. Finally,
\be
S_{K_2,K_2}(\beta)=
S_{K_1,K_1}(\beta) \langle\frac{2 i \pi}{3}-\xi\rangle\langle\frac{2 \pi}3
\rangle^2\langle \xi\rangle \sb .\ee
The analytic structure is exhibited in figure \ref{fig-12} and
\ref{fig-22}.
In both cases we showed only the direct channel poles, the crossed ones
being in a one to one correspondence. The double poles in the kink-kink
$S$-matrices can all be explained in terms of elementary scattering
processes \cite{coleman-thun}. They are
exhibited in figure \ref{fig-double}.

The $S$-matrix elements involving the breathers are the following:
\bea
S_{K_1,B_1}(\beta) &=& \langle \frac \pi 2 -\frac \xi 2\rangle_{K_1}
	      \langle\frac{5 \pi}6 -\frac \xi 2\rangle_{K_2}   \nonumber    \\
S_{K_1,B_2}(\beta)  &=& \langle \frac{2 \pi} 3\rangle^{2}
	      \langle\frac{2 \pi}3 - \xi \rangle_{K_2}
	 	\langle\xi\rangle      \nonumber \\
S_{K_2,B_1}(\beta)  &=& \langle \frac \pi 2 \rangle
	      \langle\frac{ \pi}6\rangle_{K_1}
	      \langle\frac{ \pi}6 + \xi \rangle
	      \langle-\frac{ \pi}6 + \xi \rangle   \nonumber      \\
S_{K_2,B_2}(\beta)  &=& \langle \frac \pi 3 +\frac \xi 2\rangle^{3}
      \langle \pi -\frac \xi 2\rangle
      \langle\frac{\pi}3 -\frac \xi 2\rangle_{K_1}
      \langle \pi-\frac{3 \xi} 2\rangle
      \langle-\frac{ \pi}3 +\frac{3 \xi} 2\rangle   \\
S_{B_1,B_1}(\beta)  &=& \langle  \xi \rangle
	      \langle\frac{2 \pi}3\rangle_{B_1}
	      \langle-\frac{ \pi}3 + \xi\rangle_{B_2}  \nonumber       \\
S_{B_1,B_2}(\beta)  &=& \langle \frac \pi 2 -\frac \xi 2\rangle^{2}
	      \langle-\frac{ \pi}6 +\frac{3 \xi} 2\rangle
     	      \langle-\frac{ \pi}2 +\frac{3 \xi} 2\rangle
     	      \langle\frac{ \pi}2 +\frac \xi 2\rangle
    	      \langle-\frac{ \pi}6 +\frac \xi 2\rangle_{B_1}  \nonumber       \\
S_{B_2,B_2}(\beta)  &=&(\langle\frac{2 \pi}3\rangle^3)_{B_2}
	      \langle\xi\rangle^3
		 \langle- \frac \pi 3 +\xi\rangle^2
		\langle\frac \pi 3+\xi\rangle
		\langle-\frac\pi 3+2 \xi\rangle
		\langle-\frac{2 \pi}3 +2 \xi\rangle \nonumber
\eea
The lower indices indicate the bound state corresponding to that pole.

The model $M_{5,6}$ exhibits two more breathers, even though no more kinks
are generated. This can be seen from fig \ref{fig-12} and \ref{fig-22}.
Since $\xi \le \frac\pi 2$ new breathers are created. But since $\xi \ge
\frac\pi 3$ no new kink poles come into the physical strip.
The third breather
with the mass
$$M_3=
4 m \sin \frac 5{21}\pi \sin \frac 37\pi $$
is a bound state of $K_1$ and $K_2$ at the pole
$u_{K_1,K_2}^{B_3}=\frac 2 7 \pi$. The heaviest breather
is a bound state of $K_2$ and $K_2$ at rapidity $u_{K_2,K_2}^{B_4} =
\frac 1 {21} \pi$ and has mass
$$ M_4 = 4 m \cos \frac {2 \pi} {21} \cos \frac \pi{42} \sb .$$
In \cite{marcio} the truncation method
was performed for this model, and
the all but the heaviest particle were found. This is not surprising since
for heavy particles the finiteness of the basis in the truncated Hilbert
space causes rather big systematical errors.
The remaining breather part of the $S$-matrix of this model is
\be
\begin{array}{ll}
S_{13} = \langle\frac{31}{42}\rangle_2\langle\frac 3{14}\rangle_4\langle
\frac{13}{14}\rangle_1
\frac{11}{42}\rangle\langle\frac{19}{42}\rangle
\langle\frac{17}{42}\rangle^2\sa,  &
S_{14} =\langle\frac 67\rangle_3 \langle\frac 23
\rangle^2 \langle\frac{10}{21}\rangle^2 \langle\frac 4{21}\rangle
\langle\frac 27\rangle \langle\frac 5{21}\rangle \sa,\\
& \\
S_{23} = \langle\frac 56\rangle_1 \langle\frac 5{14}\rangle^2
\langle\frac{13}{42}\rangle^2
\langle\frac 12\rangle \langle\frac 3{14}\rangle\langle\frac
{19}{42}\rangle\sa, &
S _{24} = \langle\frac{19}{21}\rangle_2
\langle\frac 27\rangle^2 \langle\frac{5}{21}\rangle\langle\frac 8{21}\rangle^2
\langle\frac 37\rangle^3 \langle\frac 17\rangle\langle\frac{10}{21}\rangle
\sa, \\
& \\
S_{33} = \langle\frac 23\rangle_3^3\langle\frac 17\rangle^2\langle
\frac {10}{21}\rangle^3
\langle\frac 27\rangle\langle\frac 8{21}\rangle\langle\frac 4{21}
\rangle\sa, & \\
& \\
S_{34}= \langle\frac {13}{14}\rangle_1 \langle\frac  {17}{42}\rangle^4
 \langle\frac 3{14}\rangle^2
\langle\frac {19}{42}\rangle^3 \langle\frac {11}{42}\rangle^3\langle
\frac 5{14}\rangle\langle\frac 5{42}\rangle\langle\frac
5{21}\rangle\sa,  &  \\
 &  \\
S_{44} = \langle\frac 23\rangle_4^5\langle\frac 27\rangle^2\langle
\frac 17\rangle^2 \langle\frac {10}{21}\rangle^5
\langle\frac 8{21}\rangle^3 \langle\frac 4{21}
\rangle^3 \langle\frac 37\rangle\langle\frac 1{21}\rangle\langle
\frac 5{21}\rangle\sb.
 &  \\
 & \end{array}\nonumber \ee
Herein the indices of the $S$-matrix elements correspond to breathers.
This is the complete breather-part of this $S$-matrix.

A final confirmation of these $S$-matrices is expected from the
thermodynamic Bethe ansatz. It involves higher level Bethe ansatz techniques,
and gets rather complicated since
there the spectrum consists of {\em two} degenerate particles.

\section{Conclusions}

We have analyzed the IRF structure which lies under $\phi_{1,2}$-perturbed
conformal field theories. They can be built as graph-state models, but
not in the usual Hecke-algebra sense but on a BWM-algebra. The advantage of
this construction is that it avoids using the vector-representation, which
leads to non-unitary scattering matrices \cite{smirnov-12}.

The disadvantage of this approach lies in the fact, that there is still one
step which essentially requires guess-work. There is no well-defined
mechanism in order to get the braid group generators fulfilling the
BWM algebra given the TLA. This is of course, because the
TLA-generators are quadratic functions of the braid group generators. Since
they do not have inverses the resolution of this quadratic relation is
highly non-trivial. If one can succeed in this point, and find the
constraints on the TLA generators, such that they give rise to a braid group
satisfying the BWM-algebra, one will have a means to define general
BWM-graph-state models. Work on this problem is in progress.

We have used this construction in order to compare the corresponding
$R$-matrix with the scattering matrices described by Smirnov. We found
that a whole series of unitary scattering matrices corresponds to one
Yang-Baxter geometry. The difference between the corresponding models
is the relation of the spectral parameter of the $R$-matrix to the
rapidity variable, and the scalar prefactors, which differ by so-called
CDD factors from each other.

The BWM geometry plays a fundamental role in the ultraviolet limit. One
finds that the asymptotic phase-shifts are in relation to the dimensions
appearing in the operator product expansion of certain fields of the
underlying CFT.

Having an explicit expression of the residues in form of $6j$ symbols, we
have rewritten the bootstrap-equations in a form, which is easier to apply.
We then used that to calculate the $S$-matrix of the higher kink, which
appears in the unitary series $M_{r,r+1} +\phi_{1,2}$.
A non-trivial degeneracy structure persists for the models $r \ge 5$.
 Using the principle
that breathers are supposed to be bound states of kinks, we find the whole
$S$-matrices involving kinks and breathers of these theories.
For $r=5$ the theory has 4 breathers and for $r \ge 6$ only 2.
The $S$-matrix elements among kinks exhibit double poles which can all be
described by elementary scattering processes of the lightest kink and the
lightest breather.

A formidable open problem is to apply the bootstrap to the $S$-matrices of
non-unitary minimal models perturbed by the operator $\phi_{1,2}$. In that
case we have seen that the $S$-matrix is unitary for the models
$\M_{r,mr\pm 1}$. These models exhibit a much more complicated bound-state
structure of kinks and breathers.
\vspace{2mm}

\noindent
{\bf Acknowledgements:}
I am grateful to M. Martins and G.Mussardo and A. Schwimmer for
many fruitful discussions on the
subject. Also thanks to Mr. Hawle and Mr. Dr\"oxler for their personal
support in burocratic issues.

\appsection

Here we collected some necessary information on the projectors.
The $R$-matrix (\ref{r-a22}) can be written in terms of projectors as
\bea
R(x) &=& (q^2 x^{-\frac 12}-x^{\frac 12} q^{-2})(q^{-3}x^{-\frac 12}
+q^3 x^{\frac 12}) P_0
+(q^2 x^{\frac 12}-x^{-\frac 12} q^{-2})(q^{-3}x^{\frac 12}+q^3
x^{-\frac 12}) P_1
\nonumber \\
&+&
(q^2 x^{-\frac 12}-x^{\frac 12} q^{-2})(q^{-3}x^{\frac 12}+q^3
x^{-\frac 12}) P_2 \sb .
\eea
We look for points where $R(x) \sim P_i$, and therefore the other terms must
vanish.
We find:
\be
\begin{array}{llll}
P_0 {\rm \,\,vanishs \,\, if:} & x=q^4 &{\rm or}& x=iq^{-6} \sa,\\
P_1                        & x=q^{-4} & & x=i q^6 \sa,\\
P_2			&x=q^4 & & x=iq^6 \sb .
\end{array}
\ee
This means that $R \sim P_0$ at $x=i q^6$ and $R\sim P_1$ at $x=q^4$.
Note that $R$ never becomes proportional to $P_2$, and therefore we can
not form bound states of spin 2 in an hypothetical $S$-matrix based on the
$R$-matrix (\ref{r-a22}).

We give now the $6j$ symbols which are necessary to carry out the
bootstrap. Note that they correspond to the fusion coefficient, graphically
displayed as
\be
\begin{picture}(70,40)(-30,0)
\thicklines
\put(0,0){\line(1,1){25}}
\put(0,0){\line(-1,1){25}}
\put(0,0){\line(0,-1){25}}
\put(-20,-10){$j_2$}
\put(15,-10){$j$}
\put(-3,15){$j_{12}$}
\put(-30,30){$j_1$}
\put(23,30){$j_3$}
\put(-5,-35){$j_{23}$}
\end{picture}
\goto
 \left \{ \begin{array}{ccc} j_{3} & j_{2} & j_{23} \\ j_{1}
&j & j_{12}  \end{array} \right \}_q \sa .
\ee

\vspace{3mm}

\noindent
The only non-zero fusion coefficients for spin 1 are
\bea
 \left \{ \begin{array}{ccc}
1 & 1 &1 \\
l &l &l       \end{array} \right \}_q
=f_{l,l,l}
&=&
(q^{2l+1}+q^{-2l-1})
\left (\frac{[2]}{[4][2l][2l+2]}\right )^{\frac 12} \sb,\nonumber\\
 \left \{ \begin{array}{ccc}
 1 & 1 & 1 \\
l &l-1&l      \end{array} \right \}_q
=f_{l,l,l-1}
&=&
\left (\frac{[2][2l+2]}{[4][2l]}\right )^{\frac 12} \sb,\nonumber\\
 \left \{ \begin{array}{ccc}
1 & 1 & 1 \\
l &l+1 &l       \end{array} \right \}_q
=f_{l,l,l+1}
&=&
-\left (\frac{[2][2l]}{[4][2l+2]}\right )^{\frac 12} \sb,\nonumber\\
 \left \{ \begin{array}{ccc}
 1 & 1 & 1 \\
l &l&l-1      \end{array} \right \}_q
=f_{l,l-1,l}
&=&
-\left (\frac{[2l-1][2][2l+2]}{[2l+3][4][2l]}\right )^{\frac 12}
\sb, \nonumber\\
 \left \{ \begin{array}{ccc}
1 & 1 & 1 \\
l &l &l+1       \end{array} \right \}_q
=f_{l,l+1,l}
&=&
\left (\frac{[2l+3][2][2l]}{[2l+1][4][2l+2]}\right )^{\frac 12}
\sb .\nonumber\eea

\noindent
The $6j$-symbols for spin $0$, that is the fusion coefficients for the
breathers, are

\bea
 \left \{ \begin{array}{ccc}
1 & 1 &0 \\
l &l &l       \end{array} \right \}_q
&=&
-\left (\frac{1}{[3]}\right )^{\frac 12} \sb,\nonumber\\
 \left \{ \begin{array}{ccc}
 1 & 1 & 0 \\
l &l&l+1      \end{array} \right \}_q
&=&
\left (\frac{[2l+3]}{[3][2l+1]}\right )^{\frac 12}\sb, \nonumber\\
 \left \{ \begin{array}{ccc}
1 & 1 & 0 \\
l &l+1 &l       \end{array} \right \}_q
&=&
\left (\frac{[2l-1]}{[3][2l+1]}\right )^{\frac 12} \sb .\nonumber\eea

%BIBLIO

\newpage

%%%%%%%%%%%%%%%%%%%%%%%%%%%%%%%%%%%%%%%%%%
%%%%%%%%%%%%%%%%%%%%%%%%%%%%%%%%%%%%%%%%%%
\begin{figure}
$$
 \begin{picture}(450,120)(0,-100)
\thicklines
\put(0,0){\line(1,0){400}}
\multiput(100,10)(40,0){6}{\line(0,-1){20}}
\multiput(100,15)(40,0){6}{\makebox(0,0){x}}
\put(260,25){\makebox(0,0){o}}
\put(220,25){\makebox(0,0){x}}
\put(220,35){\makebox(0,0){x}}
\put(180,25){\makebox(0,0){x}}
\put(300,25){\makebox(0,0){o}}
\put(100,-25){\makebox(0,0){$-3 i \xi$}}
\put(140,-25){\makebox(0,0){$-2 i \xi$}}
\put(180,-25){\makebox(0,0){$- i \xi$}}
\put(220,-55){\makebox(0,0)[t]{\shortstack{Kink\\2}}}
\put(180,-55){\makebox(0,0)[t]{\shortstack{Double\\pole}}}
\put(220,-25){\makebox(0,0){$\frac{2i \pi}3$}}
\put(260,-25){\makebox(0,0){$+ i \pi$}}
\put(300,-25){\makebox(0,0){$+2 i \pi$}}
\put(180,-35){\vector(0,-1){15}}
\put(140,-45){\vector(-1,0){75}}
\put(105,-55){\makebox(0,0)[t]{Higher Kinks}}
\end{picture}
$$
$$
 \begin{picture}(450,120)(0,-100)
\thicklines
\put(0,0){\line(1,0){400}}
\multiput(100,10)(40,0){6}{\line(0,-1){20}}
\multiput(100,15)(40,0){6}{\makebox(0,0){x}}
\put(180,25){\makebox(0,0){x}}
\put(220,25){\makebox(0,0){o}}
\put(260,25){\makebox(0,0){o}}
\put(300,25){\makebox(0,0){o}}
\put(260,35){\makebox(0,0){o}}
\put(300,35){\makebox(0,0){o}}
\put(100,-25){\makebox(0,0){$-3 i \xi$}}
\put(140,-25){\makebox(0,0){$-2 i \xi$}}
\put(180,-25){\makebox(0,0){$- i \xi$}}
\put(220,-25){\makebox(0,0){$i \pi$}}
\put(260,-25){\makebox(0,0){$+ i \xi$}}
\put(300,-25){\makebox(0,0){$+2 i \xi$}}
\put(180,-35){\vector(0,-1){15}}
\put(180,-55){\makebox(0,0)[t]{\shortstack{Double\\pole}}}
\put(140,-45){\vector(-1,0){75}}
\put(105,-55){\makebox(0,0)[t]{Higher Breathers}}
\end{picture}
$$
\caption{Pole structure of the $S$-matrix $S_{K_2,K_2}$ }
\label{fig-22}
\end{figure}
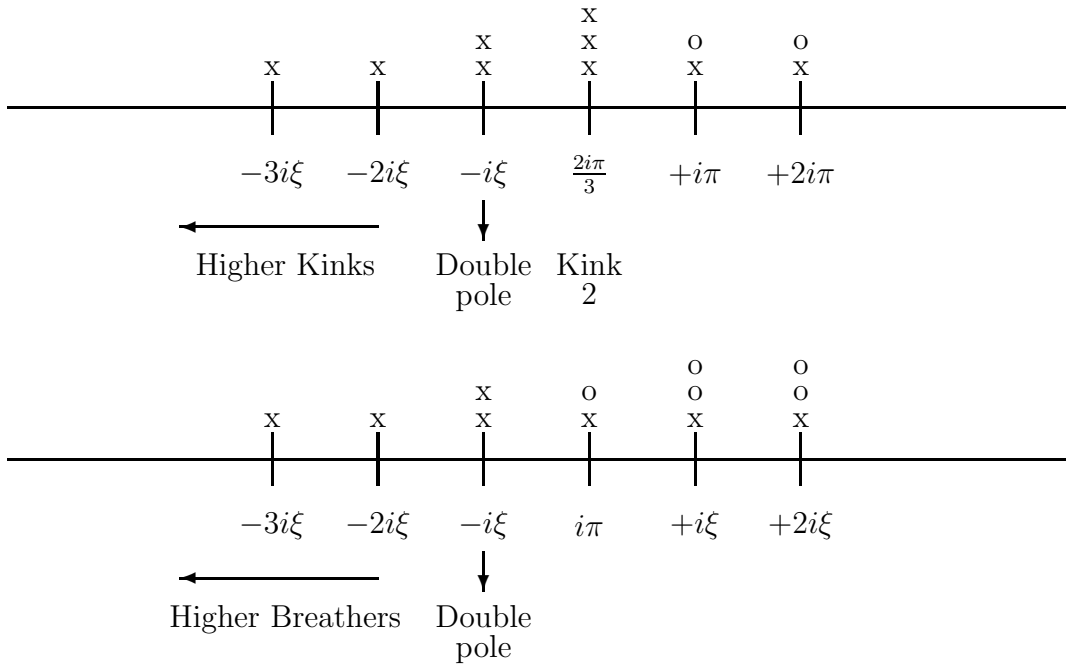

\end{document}